\documentclass[12pt]{article}
\newcommand{\beq}{\begin{equation}}
\newcommand{\eeq}{\end{equation}}

\def\beqa{\begin{eqnarray}}
\def\enqa{\end{eqnarray}}
\def\beq{\begin{equation}}
\def\enq{\end{equation}}
\def\nonum{\nonumber}

\begin{document}
\setlength{\baselineskip}{7mm}
\hspace*{0mm}

\par
\begin{center}
  {\large\bf Simplification of thermodynamic Bethe-ansatz equations }
\end{center}
\par
\begin{center}
{\bf  Minoru Takahashi}\\
{\it Institute for Solid State Physics, University of Tokyo, \\
Kashiwanoha 5-1-5, Kashiwa, Chiba, 277-8581 Japan}    \\

\vspace*{1cm}
Abstract\\
\end{center}

\par
\setlength{\baselineskip}{6.5mm}
\setlength{\parindent}{18pt}
\noindent
Thermodynamic Bethe ansatz equations for XXZ model at $|\Delta|\ge 1$ is simplified to an 
integral equation which has one unknown function. This equation is analytically 
continued to $|\Delta|<1$.\\

\noindent{\bf 1. Introduction}  \par
Thermodynamic Bethe ansatz equations for exactly solvable 
one-dimensional systems have many unknown functions.\cite{tak99} About 
the XXZ model at $|\Delta|\ge 1$, Gaudin-Takahashi equation 
contains infinite unknown functions.\cite{tak71a,gaudin71}  
Here I can simplify this 
set of equations 
to an integral equation which contains only one unknown function.
For Hamiltonian   
\beq
{\cal H}(J,\Delta,h)
=-J\sum_{l=1}^NS_l^xS_{l+1}^x+S_l^yS_{l+1}^y+\Delta (S_l^zS_{l+1}^z-{1\over 4})
-2h\sum_{l=1}S^z_l,\quad
h\ge 0,
\enq
thermodynamic Bethe ansatz equation at temperature $T$ is 
\beqa
&&\ln\eta_1(x)={2\pi J\sinh\phi \over T\phi}{\bf s}(x)+
{\bf s}*\ln(1+\eta_2(x)),\nonum\\
&&\ln\eta_j(x)={\bf s}*\ln(1+\eta_{j-1}(x))(1+\eta_{j+1}(x)),
\quad j=2,3,...,\nonum\\
&&\lim_{l\to \infty}{\ln\eta_l\over l}={2h\over T}.\label{eq:GTeq}
\enqa
Here we put
\beq
\Delta=\cosh\phi,\, Q\equiv \pi/\phi,\,
{\bf s}(x)={1\over 4}\sum_{n=-\infty}^\infty 
{\rm sech}\Bigl({\pi (x-2nQ)\over 2}\Bigr) ,\,
{\bf s}*f(x)\equiv \int^Q_{-Q}{\bf s}(x-y)f(y){\rm d}y.
\enq
The free energy per site is 
\beq
f={2\pi J\sinh \phi\over \phi}\int^Q_{-Q}{\bf a}_1(x){\bf s}(x){\rm d}x
-T\int^Q_{-Q}{\bf  s}(x)\ln(1+\eta_1(x)){\rm d}x,\,
{\bf a}_1(x)\equiv{\phi\sinh\phi/(2\pi)\over \cosh \phi -\cos(\phi x)}.
\label{eq:fen1}
\enq
On the contrary new equation is 
\beqa
&&u(x)=2\cosh({h\over T})+\oint_C {\phi\over 2}
\Bigl(\cot{\phi\over 2}[x-y-2i]\exp[-{2\pi J\sinh \phi\over T\phi}{\bf a}_1(y+i)]
\nonum\\
&&
+\cot{\phi\over 2}[x-y+2i]\exp[-{2\pi J\sinh \phi\over T\phi}
{\bf a}_1(y-i)]\Bigr){1\over u(y)}{{\rm d}y\over 2\pi i },\label{eq:neweq}
\enqa
and free energy is given by
\beq
f=-T\ln u(0).\label{eq:fen2}
\enq
Contour $C$ is an arbitrary closed loop counterclockwize around $0$. 
$2nQ, n\ne 0$ and $\pm 2i+2nQ$ should be outside of this loop. This 
loop should not contain zeros of $u(y)$. It is expected that $u(y)$ 
has not zero in region $|\Im y|\le 1$. 
This equation can be calculated even at imaginary $\phi$, 
namely, $|\Delta|<1$ case. This equation converges numerically 
at least $T/J>0.07$. The results 
coincide with those of older equations. Then this equation unifies 
Gaudin-Takahashi equation for $\Delta\ge 1$ and Takahashi-Suzuki equation 
for $|\Delta|<1$.\cite{TS72} 

\par
\noindent
{\bf 2. Derivation}\\
We should note that if 
\beq
g(x)={\bf s}*h(x),\label{eq:conv1}
\enq
it stands
\beq
g(x+i)+g(x-i)=h(x).\label{eq:conv2}
\enq
The Fourier transform of (\ref{eq:conv1}) is 
$$\tilde{g}(\omega)={1\over e^\omega+e^{-\omega}}\tilde{h}(\omega),\quad 
\omega={\pi\over Q}n.$$
Then we have 
$$(e^\omega+e^{-\omega})\tilde{g}(\omega)=\tilde{h}(\omega),$$
and obtain (\ref{eq:conv2}). \par
Then set of equations (\ref{eq:GTeq}) is rewritten as 
\beqa
&&\eta_1(x+i)\eta_1(x-i)=
\exp\Bigl[{2\pi J\sinh \phi\over T\phi}\sum_n\delta(x-2nQ)\Bigr]
(1+\eta_2(x)),\nonum\\
&&\eta_j(x+i)\eta_j(x-i)=(1+\eta_{j-1}(x))(1+\eta_{j+1}(x)),\quad 
j=2,3,...,\nonum\\
&& \lim_{l\to \infty}{\ln \eta_l\over l}={2h\over T}.\label{eq:eq7}
\enqa
At $J=0$ this set of equation becomes a difference equation and 
we have analytical solution,
$$\eta_j=\Bigl({\sinh (j+1)h/T\over \sinh h/T}\Bigr)^2-1.$$
We can expand perturbatinally as power series of $J/T$. Very surprisingly 
$\eta_j(x)$ has singularity only at $x=\pm ji,\, \pm (j+2)i +2nQ$. 
So we assume that $\eta_j(x)$ is 
univalent on the complex plane of $x$ and $1+\eta_j(x)$ is factorized as 
follows:
\beq
1+\eta_j(x)=A_j(x-ji)\overline{A_j}(x+ji)B_j(x-(j+2)i)\overline{B_j}(x+(j+2)i).
\label{eq:eta+1}
\enq
Here functions $A_j(x),\, B_j(x)$ are periodic with periodicity $2Q$ and 
have singularity only at $x=2nQ$ and 
$$\overline{A_j}(x)\equiv\overline{A_j(\overline{x})},\quad 
\overline{B_j}(x)\equiv\overline{B_j(\overline{x})}.$$
One should note that $\overline{A_j(x)}$ is not an analytic function of $x$ 
but $\overline{A_j}(x)$ is analytic. 
Then second equation of (\ref{eq:eq7}) becomes
\beqa
&&\eta_j(x+i)\eta_j(x-i)=\nonum\\
&&A_{j-1}(x-(j-1)i)\overline{A_{j-1}}(x+(j-1)i)B_{j-1}(x-(j+1)i)
\overline{B_{j-1}}(x+(j+1)i)\nonum\\
&&\times A_{j+1}(x-(j+1)i)\overline{A_{j+1}}(x+(j+1)i)
B_{j+1}(x-(j+3)i)\overline{B_{j+1}}(x+(j+3)i).
\enqa
If we assume that $\eta_j(x)$ is factorized as 
$$
\eta_j(x)=X(x-ji)Y(x+ji)Z(x-(j+2)i)W(x+(j+2)i),
$$
$\eta_j(x+i)\eta_j(x-i)$ is
\beqa
&&X(x-(j-1)i)Y(x+(j-1)i)X(x-(j+1)i)Y(x+(j+1)i)\nonum\\
&&Z(x-(j+1)i)W(x+(j+1)i)Z(x-(j+3)i)W(x+(j+3)i).\nonum
\enqa
Comparing the singularity at $x=\pm(j-1)i$ and $x=\pm(j+3)i$ we have
$X\sim A_{j-1}$, $Y\sim\overline{A_{j-1}}$, $Z\sim B_{j+1}$ and 
$W\sim \overline{B_{j+1}}$. 
Then $\eta_j(x)$ should be factorized as:
\beq
\eta_j(x)=A_{j-1}(x-ji)\overline{A_{j-1}}(x+ji)B_{j+1}(x-(j+2)i)
\overline{B_{j+1}}(x+(j+2)i).
\label{eq:eta}
\enq
Considering the singularity at $x=\pm (j+1)i$ we have
\beq
{A_{j-1}(x)\over B_{j-1}(x)}={A_{j+1}(x)\over B_{j+1}(x)}.\label{eq:ratio}
\enq
Using the first equation of (\ref{eq:eq7}) and noting that 
$$\delta(x)={1\over 2\pi}\Bigl\{{-i\over x-i\epsilon}+{i\over x+i\epsilon}
\Bigr\},$$
we have
\beq
A_0(x)=\exp({J\sinh \phi\over T\phi}\sum_n{-i\over  x-2nQ-i\epsilon})
=\exp\Bigl({J\sinh \phi\over 2T i}\cot{\phi\over 2}(x-i\epsilon)\Bigr),\, 
B_0(x)=1.\label{eq:A0}
\enq 
From (\ref{eq:ratio}) we have
\beq
{A_{2j}(x)\over B_{2j}(x)}={A_0(x)\over B_0(x)}=A_0(x), \quad
{A_{2j+1}(x)\over B_{2j+1}(x)}={A_{1}(x)\over B_{1}(x)}=A_0'(x). 
\label{eq:A&A'}
\enq
One can show that $A_0(x)=A_0'(x)$. 
Consider $\eta_{2j+1}^{-1}(x+(2j+1)i)$. This quantity 
approaches $0$ in the limit of infinite $j$. Of course near 
the singularity at $2i, 0, -(4j+2)i$ and $-(4j+4)i$ the deviation from $0$ 
becomes big. Nevertheless the region on the complex ~~~~~~~plane where 
$|\eta_{2j+1}^{-1}(x+(2j+1)i)|>\epsilon$
is expected to be narrower as $j$ goes to infinity. So we have
\beqa
&& 1=\lim_{j\to \infty}{1+\eta_{2j+1}(x+(2j+1)i)\over \eta_{2j+1}(x+(2j+1)i)}
=\lim_{j\to \infty}{A_0'(x)\over A_0(x)}{\overline{A_0'}(x+(4j+2)i)\over \overline{A_0}(x+(4j+2)i)}\nonum\\
&&\times
\Bigl[{B_{2j+1}(x)\overline{B_{2j+1}}(x+(4j+2)i)\over
B_{2j}(x)\overline{B_{2j}}(x+(4j+2)i)}\cdot
{B_{2j+1}(x-2i)\overline{B_{2j+1}}(x+(4j+4)i)\over 
B_{2j+2}(x-2i)\overline{B_{2j+2}}(x+(4j+4)i)}
\Bigr].
\label{eq:1+eta-}
\enqa
The first fraction in the big bracket should go to $e^{h/T}$ and the 
second should go to $e^{-h/T}$.  
The bracket in (\ref{eq:1+eta-}) becomes $1$ in the limit of infinite $j$. 
Then we have 
$${A_0'(x)\over A_0(x)}={\overline{A_0}(i\infty)\over \overline{A_0'}(i\infty)}
=\alpha.$$
As ${\overline{A_0}(i\infty)/ \overline{A_0'}(i\infty)}=
\overline{A_0(-i\infty)/A_0'(-i\infty)}=1/\overline{\alpha}$, 
we have $|\alpha|=1$. If we choose 
the phase factor $\alpha$ is $1$, 
the ratio of $A_j(x)$ and $B_j(x)$ 
is always $A_0(x)$. 
From (\ref{eq:eta+1}) and (\ref{eq:eta}) we have
\beqa
&&A_{j-1}(x-ji)\overline{A_{j-1}}(x+ji)B_{j+1}(x-(j+2)i)
\overline{B_{j+1}}(x+(j+2)i)+1\nonum\\
&&=A_j(x-ji)\overline{A_j}(x+ji)B_j(x-(j+2)i)\overline{B_j}(x+(j+2)i).
\enqa
At $j=1$ we have
\beq
B_1(x-i)\overline{B_1}(x+i)B_1(x-3i)\overline{B_1}(x+3i)=
{1\over A_0(x-i)\overline{A_0}(x+i)}+
B_2(x-3i)\overline{B_2}(x+3i).\label{eq:BBBB}
\enq
$B_1, B_2$ are unknown functions. But this equation and condition 
\beq
\lim_{x\to i\infty}B_1(x-i)\overline{B_1}(x+i)B_1(x-3i)\overline{B_1}(x+3i)
=(2\cosh h/T)^2,
\enq
are sufficient to determine these functions. Put 
\beq
u(x)=B_1(x-2i)\overline{B_1}(x+2i).
\enq
Equation (\ref{eq:BBBB}) is written as
\beq
u(x+i)={1\over A_0(x-i)\overline{A_0}(x+i)u(x-i)}+
{B_2(x-3i)\overline{B_2}(x+3i)\over u(x-i)}.
\enq
The l.h.s. has singularity at $i, -3i$. The first term of r.h.s. has 
at $i, -i, 3i$. The second term of r.h.s has at $3i, -3i, -i$. 
Assume that $u(x)$ is expanded as follows
\beq
u(x)=2\cosh({h\over T})+\sum_{j=1}^{\infty}
\sum_n{c_j\over (x-2nQ-2i)^j}
+\sum_{j=1}^{\infty}
\sum_n{\overline{c_j}\over(x-2nQ+2i)^j}.\label{eq:cj}
\enq
Consider the contour integral around $x=i$. Coefficients $c_j$ 
is determined by
\beq
c_j=\oint {(x-i)^{j-1}\over A_0(x-i)\overline{A_0}(x+i)u(x-i)}
{{\rm d}x\over 2\pi i}
=\oint {y^{j-1}\over A_0(y)\overline{A_0}(y+2i)u(y)}{{\rm d}y\over 2\pi i}.
\enq
The first sum of r.h.s. of (\ref{eq:cj}) is
\beqa
&&\sum_{j=1}^\infty 
\oint\sum_n{\exp[-{2\pi J\sinh \phi\over T\phi}{\bf a}_1(y+i)]\over (x-2nQ-2i)^j}
{y^{j-1}\over u(y)}{{\rm d}y\over 2\pi i}
=\oint \sum_n
{\exp[-{2\pi J\sinh \phi\over T\phi}{\bf a}_1(y+i)]\over x-y-2nQ-2i}
{1\over u(y)}{{\rm d}y \over 2\pi i}\nonum\\
&&=\oint {\phi\over 2}\cot{\phi\over 2}(x-y-2i)
\exp[-{2\pi J\sinh \phi\over T\phi}{\bf a}_1(y+i)]
{1\over u(y)}{{\rm d}y\over 2\pi i}. 
\enqa
The second sum is calculated in similar way. Thus we get (\ref{eq:neweq}).
From equation (\ref{eq:eta+1}) we have
\beq
1+\eta_1(x)=
\exp\Bigl({2\pi J\sinh\phi\over T\phi}{\bf a}_1(x)\Bigr)u(x+i)u(x-i).
\enq
Substituting this into eq.(\ref{eq:fen1}) we get eq.(\ref{eq:fen2})
\beq
f=-T\int^Q_{-Q}{\bf s}(x)[\ln u(x+i)+\ln u(x-i)]{\rm d}x=-T\ln u(0).
\enq
\\
{\bf 3. Analytical solutions}\\
{\it Ising limit}\\
In this limit we put $\Delta\to \infty, J=J_z/\Delta$. As $\phi$  goes to 
$\infty$, $2Q$¡¡~~becomes $0$. Then function  ${2\pi J\sinh \phi\over T\phi}{\bf a}_1(y)$ 
is $0$ at $|\Im y|>1$ and $J_z/T$ at $ |\Im y|<1$. Function $u(y)$ is $2\cosh h/T$ at 
$|\Im y|>2$ and also a constant $u(0)$ at $|\Im y|<2$. From eq.(\ref{eq:neweq}) 
we have
\beq
u(0)=2\cosh(h/T)+{1-\exp(-J_z/T) \over u(0)}.
\enq
So we have 
$u(0)=\cosh (h/T)+\sqrt{\sinh^2 (h/T)+\exp(-J_z/T)}$ 
and known free energy of the Ising model.\\

Kuniba, Sakai and Suzuki \cite{kuni98} showed that Takahashi-Suzuki equation 
for $|\Delta|<1, ~ h=0$ can be derived from the quantum transfer matrix 
and its fusion hierarchy matrices. We are confident to derive 
eqs.(\ref{eq:neweq}, \ref{eq:fen2}) from the quantum transfer matrix and 
its fusion hierarchies. Details will be published elsewhere. \\

This research was
supported in part by Grants-in-Aid for the Scientific Research  (B) No. 11440103
from the Ministry of Education, Science and Culture, Japan.

\end{document}